\documentstyle[aps,prl,preprint]{revtex} 
\begin{document} 
\draft 
\title{On the origin of the irreversibility line in thin $YBa_2Cu_3O_{7-\delta }$ 
films with and without columnar defects} 
\author{R.~Prozorov$^1$, M.~Konczykowski$^2$, B.~Schmidt$^2$, Y.~Yeshurun$^1$, A.~ 
Shaulov$^1$, C.~Villard$^3$\thanks{\protect\underline{Permanent address:} 
Laboratoire EPM Matformag, CNRS BP166, 38042 Grenoble Cedex, France\noindent}%
, G.~Koren$^3$} 
\address{$^1$Institute of Superconductivity, Department of Physics, Bar-Ilan\\ 
University, 52900 Ramat-Gan, Israel\\ 
$^2$Laboratoire des Solides Irradi\'{e}s, Ecole Polytechnique, 91128\\ 
Palaiseau, France\\ 
$^3$Department of Physics, Technion, Haifa} 
\date{June 10, 1996} 
\maketitle 
 
\begin{abstract} 
We report on measurements of the angular dependence of the irreversibility 
temperature $T_{irr}\left( \theta \right) $ in $YBa_2Cu_3O_{7-\delta }$ thin 
films, defined by the onset of a third harmonic signal and measured by a 
miniature Hall probe. From the functional form of $T_{irr}\left( \theta 
\right) $ we conclude that the origin of the irreversibility line in 
unirradiated films is a dynamic crossover from an unpinned to a pinned 
vortex liquid. In irradiated films the irreversibility temperature is 
determined by the trapping angle. 
\end{abstract} 
 
 
\section{Introduction} 
 
The origin of the irreversibility line (IRL) in the field-temperature ($H-T$%
) phase diagram of high-temperature superconductors (HTS) is intriguing and 
still a widely discussed topic \cite 
{muller,yeshurun,brandt,brandt2,blatter,matsushita,majer,zeldov0,zeldov}. 
Experimentally, this line is defined as the border line at which the 
magnetic response of the sample changes from irreversible to reversible. In 
HTS, large fluctuations and relatively weak pinning lead to a rich $H-T$ 
phase diagram with a variety of dynamic and static transitions which can be 
responsible for the appearance of magnetic reversibility \cite 
{brandt,brandt2,blatter,vinokur,feigelman2,geshkenbein}. Thus, a thorough 
experimental investigation of the IRL is important for the understanding of 
the vortex-lattice behavior in superconductors in general and of the 
mechanisms responsible for the onset of irreversible magnetic response, in 
particular. 
 
Several models, like thermally activated depinning \cite 
{yeshurun,brandt,matsushita,palstra,neminsky}, vortex-lattice melting \cite 
{worthington,houghton,feigelman,andrade,hyun,beck,kwok} and a transition 
from vortex glass to vortex fluid \cite{fisher,deak,giura}, were proposed to 
identify the origin of the IRL in HTS. Also, attention was given to the 
possibility of pinning in the vortex-liquid phase \cite 
{blatter,vinokur,feigelman2,obara} and to different dissipation mechanisms 
above the melting line \cite{blatter,palstra,qui,chien,yeh}. Irreversibility 
due to geometrical \cite{majer,zeldov} or surface barriers \cite{sb} have 
also been proposed, but these mechanisms are less probable in thin $%
YBa_2Cu_3O_{7-\delta }$ films with strong pinning. The irreversibility line 
may be affected by sample-dependent properties such as the nature and 
density of pinning centers and by intrinsic or extrinsic anisotropy. For 
example, in superconductors with columnar defects, the irreversibility line 
may either be identified with the Bose-glass transition \cite 
{blatter,yeh,fisher2,beek,jiang,larkin,budhani,konczykowski,ruyter,krusin}, 
or related to the concept of a trapping angle \cite{zech}. The configuration 
of the columnar defects is also very important, since it affects possibility 
for different types of depinning mechanism. A splayed configuration, for 
example, inhibits creep from columnar defects \cite{hwa,civale}. Similarly 
''crossed'' defects (i.e. defects at angles $\pm \theta $) were shown to act 
collectively, i.e., they introduce unidirectional anisotropy such that the 
current density reaches its maximum for magnetic field directed in a mid 
angle between defects \cite{schuster,CollectiveAction}. 
 
Experimentally, the situation is even more complex, since different 
techniques (magnetization loops, field-cool vs. zero-field-cool DC 
magnetization, peak in the imaginary part of the first harmonic etc.) yield 
different IRLs \cite{deak,yacobi}. To a great extent, the reliability of the 
determination of IRL depends on the criterion for the onset of the 
irreversibility. We determine the irreversibility temperature at given DC 
field by the onset of third harmonic in the $ac$ response, which, we 
believe, is one of the most reliable methods for contactless determination 
of the IRL \cite{tirr-V3}. In most experiments $T_{irr}$ is measured as a 
function of the external field $H$. This information is insufficient to 
distinguish between different models for the origin of the irreversibility. 
Additional information, like the frequency dependence of the IRL \cite 
{deak,prozorov} or its angular variation \cite 
{andrade,beck,kwok,jiang,budhani,ruyter,iye}, is needed. 
 
In this paper we report on a study of the angular dependence of the 
irreversibility temperature $T_{irr}\left( \theta \right) $ in thin $%
YBa_2Cu_3O_{7-\delta }$ ($YBCO$) films before and after irradiation with Pb 
ions. 
 
\section{Experimental} 
 
The $1500\ \AA $ $YBCO$ films were ''sandwiched'' between $SrTiO_3$ layers  
\cite{koren}. First, a $500\ \AA $ layer of $SrTiO_3$ was deposited on a $%
MgO $ substrate. Then, the $YBCO$ film was laser ablated on top of the $%
SrTiO_3$ and finally, the $YBCO$ was covered by a protective $300\ \AA $ 
layer of $SrTiO_3$. All three samples have the same lateral dimensions of $%
100\times 500\ \mu m^2$. One film, denoted as ${\em REF}$, was used as a 
reference sample. The other two, {\em UIR }and{\em \ CIR}, were irradiated 
at GANIL with $2\times 10^{11}\ ions/cm^2$ $5.8\ GeV$ Pb ions along the $c-$%
axis and along $\theta =\pm 45^o$, respectively. ({\em UIR }and{\em \ CIR }%
stand for ''uniform irradiation'' and ''crossed irradiation'', 
respectively). The superconducting transition temperatures, measured by a  
{\em Quantum Design SQUID }susceptometer and defined as the onset of the 
Meissner expulsion in a DC field of $5\;G$, are $T_c\approx 89\ K$ for the 
samples {\em REF} and {\em UIR} and $88\ K$ for {\em CIR}. 
 
For the $ac$ measurements reported below we used a miniature $80\times 80$\ $%
\mu m^2$ InSb Hall-probe, which was positioned in the center of the sample. 
The $1\ G$ $ac$ magnetic field, always parallel to the $c-$ axis, was 
induced by a small coil surrounding the sample. An external $dc$ magnetic 
field, up to $H_a=1.5\ Tesla$, could be applied at any direction $\theta $ 
with respect to the $c-$ axis. In our experiments ${\em DC}$ magnetic field 
was always turned on at a fixed angle at $T>T_c$ and then the $ac$ response 
was recorded during sample cooling. The irreversibility temperature, $%
T_{irr}\left( \theta \right) $ is defined as the onset of the third harmonic 
signal in the $ac$ response measured by the Hall probe \cite{tirr-V3}. This 
procedure was repeated for various $dc$ fields and at various angles $\theta  
$ of the field with respect to the $c-$axis. 
 
\section{Results} 
 
Fig. \ref{RawData} presents measurements of $V_3$, the third harmonic in the  
{\it ac} response, versus temperature $T$, during field-cooling at $1\ Tesla$ 
for the sample {\em REF} at various angles between $0$ and $90^o$. 
Apparently, as the angle $\theta $ increases the whole $V_3$ curve shifts to 
higher temperatures and becomes narrower. The onset of irreversibility, $%
T_{irr}\left( \theta \right) $, is defined by the criterion $%
V_3^{onset}=0.05 $ in the units of Fig. \ref{RawData}. 
 
Fig. \ref{refref} exhibits typical $T_{irr}\left( \theta \right) $ data for 
the unirradiated sample {\em REF}, measured at two values of the external 
field: $0.5\ Tesla$ and $1\ Tesla$. Both curves exhibit a shallow minimum 
around $\theta =0$ and they reach their maximum values for $H$ along the $ab$ 
plane, at angles $\theta =\pm 90^o$. We also measured the frequency 
dependence of $T_{irr}$ for the same values of magnetic field. As shown in 
Fig. \ref{reffr}, the slope $\partial T_{irr}/\partial \ln \left( f\right) $ 
is larger for larger field. 
 
The sample irradiated along the $c-$axis exhibits additional feature - a  
{\em peak }around{\em \ }$\theta =0$. This is clearly shown in Fig. \ref 
{refuir} where we compare $T_{irr}\left( \theta \right) $ at $H=1\ Tesla$ 
for the samples {\em REF} and {\em UIR}. As discussed below, this peak is a 
signature of the unidirectional magnetic anisotropy induced by the columnar 
defects. Intuitively, one would therefore expect two peaks, along $\theta 
=\pm 45^o$, for the third sample, {\em CIR}, crossed-irradiated at $\theta 
=\pm 45^o$. Instead, we find one strong peak around $\theta =0$, similar to 
that found in BSCCO crystals \cite{CollectiveAction}. This is demonstrated 
in Fig. \ref{refcir} where we compare $T_{irr}\left( \theta \right) $ at $%
H=0.5\ Tesla$ for this sample ({\em CIR}) and for the unirradiated sample (%
{\em REF}). We argue below that the peak around $\theta =0^o$ is a result of 
a collective action of the crossed columnar defects, and its origin is the 
same as that for unidirectional enhancement of critical current density 
observed in {\em BSCCO }crystals \cite{schuster,CollectiveAction}. 
 
\section{Analysis} 
 
The ''true'' irreversibility temperature $T_0$ is defined as a temperature 
above which the pinning vanishes. Such disappearance of pinning can be of 
static (true phase transition), as well as of dynamic origin (gradual 
freezing, pinning in liquid). In practice, one determines the 
irreversibility temperature $T_{irr}\left( \Delta \right) $ as the 
temperature above which the critical current density is less than some 
threshold value $\Delta $. Therefore, by definition, $T_0=\lim\limits_{%
\Delta \rightarrow 0}\left( T_{irr}\left( \Delta \right) \right) $. The 
apparent current depends on temperature $T$, magnetic field $B$ and the 
frequency $f$ of the exciting field which defines a characteristic 
time-scale $1/f$ for the experiment. By solving the equation $j\left( 
T,B,f\right) =\Delta $ with respect to $T$ one finds the experimental 
irreversibility temperature $T_{irr}$ for constant $B$ and $f$. In the 
following we argue that in our experiments the measured $T_{irr}$ is a good 
approximation of $T_0$. In order to estimate $T_{irr}$ we employ a general 
form for the apparent current density {\em in the vicinity of the 
irreversibility line} (IRL) \cite{brandt2,blatter,matsushita,jc(h)}:  
\begin{equation} 
j\left( T,B,f\right) \propto j_c\left( 0\right) \frac{\left( 1-T/T_0\right) 
^\alpha }{\left( B/B_0\right) ^\beta }\left( \frac f{f_0}\right) ^\gamma 
\label{j} 
\end{equation} 
where the parameters $B_0$ and $f_0$ are temperature independent (Eq. \ref{j} 
is thus valid only in a narrow temperature interval near the IRL and for 
fields larger than $H_{c1}$). From Eq. \ref{j} we get:  
\begin{equation} 
T_{irr}=T_0\left( B\right) \left( 1-\left( \frac \Delta {j_c\left( 0\right) }%
\left( \frac B{B_0}\right) ^\beta \left( \frac{f_0}f\right) ^\gamma \right) 
^{\frac 1\alpha }\right)  \label{tirr} 
\end{equation} 
Inserting reasonable numerical estimates: $j_c\left( 0\right) \simeq 10^7\ 
A/cm^2$, $\Delta \simeq 100\ A/cm^2$ for our experimental resolution, $%
B_0\simeq 10^3\ G$, $B\simeq 10^4\ G$, $\beta \simeq 1$ \cite{jc(h)}, $%
\gamma \simeq 1$ \cite{matsushita}, $f\simeq 10^2\ Hz$, and $f_0\simeq 10^7\ 
Hz$ \cite{matsushita}, we get from Eq. \ref{tirr}: $T_{irr}=T_0\left( 
B\right) \left( 1-0.005^{1/\alpha }\right) $. Thus, with $0.5\%$ accuracy we 
may say that $T_{irr}$, the measured onset of the third harmonic component 
in the $ac$ response, marks some ''true'' irreversibility crossover line $%
T_0\left( B\right) $. The nature of this line $T_0\left( B\right) $ is our 
main interest, since, as discussed in the introduction, it is directly 
related to the pinning properties of vortex lattice in type-II 
superconductors at high temperatures. 
 
\subsection{Unirradiated YBCO film} 
 
We turn now to consider the effect of the intrinsic anisotropy on $%
T_{irr}\left( \theta \right) $. Following the anisotropic scaling approach 
proposed by Blatter et al.\cite{blatter2,comment}, we replace $T$ by $%
\varepsilon T$ and $B$ by $B_{eff}=\varepsilon _\theta B$, where $%
\varepsilon _\theta =\sqrt{\cos ^2\left( \theta \right) +\varepsilon ^2\sin 
^2\left( \theta \right) }$ and $\varepsilon \approx 1/7$ is the anisotropy 
parameter for $YBCO$. It should be emphasized that we can use this scheme 
only in the case of {\em intrinsic} anisotropy $\varepsilon =\sqrt{m_{ab}/m_c%
}$, where $m_c$ and $m_{ab}$ denote the effective masses of the electron 
along the $c-$axis and in the $ab-$ plane, respectively. In the case of some  
{\em extrinsic} magnetic anisotropy, (columnar defects or twin planes), the 
critical current depends on the angle not only via the effective magnetic 
field $B_{eff}$, but also because of this extrinsic anisotropy. 
 
As we have already indicated in the Introduction, there are several possible 
origins for a crossover from irreversible to reversible magnetic behavior in 
unirradiated samples. We exclude the vortex-glass to vortex fluid transition 
as a possible origin for the IRL, because this transition was shown to occur 
at temperatures lower than the onset of dissipation \cite{deak,obara,kotzler}%
. The thermal depinning temperature {\em increases} with increase of field  
\cite{blatter} $T_{dp}\propto \sqrt{B}$ and, therefore is excluded as well. 
Vortex-lattice melting transition is believed to be responsible for the 
appearance of reversibility \cite{worthington,houghton,andrade,hyun}. The 
explicit angular dependence of $T_m$ was derived by Blatter et al. \cite 
{blatter,blatter2} using their scaling approach:  
\begin{equation} 
T_m\left( \theta \right) \simeq 2\sqrt{\pi }\varepsilon \varepsilon 
_0c_L^2\left( \Phi _0/B\varepsilon _\theta \right) ^{1/2}\approx \frac{%
c_L^2T_c}{\sqrt{\beta _mGi}}\left( 1-\frac{T_m}{T_c}\right) \left( \frac{%
H_{c2}\left( 0\right) }{\varepsilon _\theta B}\right) ^{\frac 12} 
\label{tm0} 
\end{equation} 
where $\Phi _0$ is the flux quantum, $\xi $ is the coherence length, $\beta 
_m\approx 5.6$ is a numerical factor, estimated in \cite{blatter}, $%
c_L\simeq 0.1$ is the Lindemann number, $Gi=\left( T_c/\varepsilon 
H_{c2}\left( 0\right) \xi ^3\left( 0\right) \right) ^2/2$ is the Ginzburg 
number, and $H_{c2}\left( 0\right) $ is the {\em linear} extrapolation of 
the upper critical field from $T_c$ to zero. Solving Eq. \ref{tm0} with 
respect to $T_m$ we get:  
\begin{equation} 
T_m\left( \theta \right) \simeq \frac{T_c}{1+\left( \frac{\beta _mGi}{%
c_L^4H_{c2}\left( 0\right) }\right) ^{\frac 12}\left( \varepsilon _\theta 
B\right) ^{\frac 12}}\equiv \frac{T_c}{1+C\sqrt{\varepsilon _\theta B}}. 
\label{tm} 
\end{equation} 
 
\noindent Eq. \ref{tm} predicts that the melting temperature decreases as $%
B_{eff}$ increases. This is due to the fact that the inter-vortex distance $%
a_0^2\propto 1/B_{eff}$ decreases faster than the characteristic amplitude 
of fluctuations $\left\langle u^2\left( B_{eff},T_m\right) \right\rangle 
_{th}\propto 1/\sqrt{B_{eff}}$. Therefore, the condition for the 
vortex-lattice melting $\left\langle u^2\left( B_{eff},T_m\right) 
\right\rangle _{th}\simeq c_L^2a_0^2$ implies larger melting temperatures 
for smaller effective fields, i. e., for larger angles. In agreement with 
this prediction, the experimental data of Fig. \ref{refref} show that $%
T_{irr}$ increases with the angle. i.e., decreases with $B_{eff}$. The solid 
lines in Fig. \ref{refref} are fits to equation \ref{tm}. From this fit we 
get $C\simeq 0.0005$. However, a reasonable estimate of $C\simeq \sqrt{%
\left( \beta _mGi/c_L^4H_{c2}\left( 0\right) \right) }$ yields $C\simeq 0.01$%
, where we take $H_{c2}\left( 0\right) =5\cdot 10^6\ G$, $c_L=0.1$, $Gi=0.01$%
, and $\beta _m=5.6$ \cite{blatter}. Also, Yeh et al. showed that the onset 
of irreversibility occurs above the melting temperature (Ref.\cite{yeh}, 
Fig. 4). In addition, the important effect of the frequency (see Fig. \ref 
{reffr}) is not included in Eq. \ref{tm}. 
 
We discuss now another possibility for the onset of the irreversibility, 
namely, pinning in the vortex liquid (for a discussion see Ch. VI in Blatter 
et al. \cite{blatter} and references therein). Any fluctuation in the vortex 
structure in the liquid state has to be averaged over the characteristic 
time scale for pinning $t_{pin}$. In the absence of viscosity the only 
fluctuations in the liquid state are thermal fluctuations, which have a 
characteristic time $t_{th}<<t_{pin}$. (As shown in \cite{blatter} $%
t_{pin}/t_{th}\propto j_0/j_c$, where $j_0$ is the depairing current). Thus, 
such a liquid is always unpinned. The situation is different for a liquid 
with finite viscosity. In this case there exists another type of excitations 
in the vortex structure, i.e. {\em plastic} deformations with a 
characteristic time scale $t_{pl}$. The energy barrier, corresponding to 
plastic deformation is shown to be \cite{blatter,geshkenbein}:  
\begin{equation} 
U_{pl}\simeq \gamma \varepsilon \varepsilon _0a_0\simeq \gamma \left( \frac{%
H_{c2}}{4Gi}\right) ^{\frac 12}\left( T_c-T\right) B^{-1/2}  \label{upl} 
\end{equation} 
where $\gamma $ is a coefficient of the order of unity. The corresponding 
characteristic time scale is:  
\begin{equation} 
t_{pl}\sim t_{th}\exp \left( U_{pl}/T\right)  \label{tpl} 
\end{equation} 
Thus, depending on the viscosity, $t_{pl}$ can be smaller or larger than $%
t_{pin}$. In the latter case, after averaging over a time $t_{pin}$, the 
vortex structure remains distorted and such a liquid shows irreversible 
magnetic behavior. Thus, on the time scale of $t_{pin}$ the distorted vortex 
structure is pinned. The crossover between pinned and unpinned liquid occurs 
at temperature $T_k$ where the characteristic relaxation time for pinning $%
t_{pin}\left( T\right) $ becomes comparable to that for plastic motion $%
t_{pl}\left( T\right) $. Thus, using Eqs. \ref{upl} and \ref{tpl} we obtain:  
\begin{equation} 
T_k=\frac{T_c}{1+\frac 1\gamma \left( \frac{4Gi}{H_{c2}\left( 0\right) }%
\right) ^{\frac 12}\ln \left( t_{pin}/t_{th}\right) \sqrt{B}}  \label{tkb} 
\end{equation} 
Finally, using the anisotropic scaling \cite{blatter2} we may rewrite Eq.  
\ref{tkb} for $f_{pin}<f<f_{th}$ as:  
\begin{equation} 
T_{irr}\left( \theta \right) =T_k\left( \theta \right) =\frac{T_c}{1+\frac 1%
\gamma \left( \frac{4Gi}{H_{c2}\left( 0\right) }\right) ^{\frac 12}\ln 
\left( f_{th}/f\right) \sqrt{B}\sqrt{\varepsilon _\theta }}\equiv \frac{T_c}{%
1+A\sqrt{\varepsilon _\theta B}}  \label{Tk} 
\end{equation} 
with $f_{th}\equiv 1/t_{th}$ and $f_{pin}\equiv 1/t_{pin}$. Note the 
apparent similarity with the expression for the melting temperature, Eq. \ref 
{tm}. The numerical estimate for $A=\frac 1\gamma \left( \frac{4Gi}{%
H_{c2}\left( 0\right) }\right) ^{\frac 12}\ln \left( f_{th}/f\right) $ 
gives: $A\simeq 10^{-4}\ln \left( f_{th}/f\right) /\gamma $. This is in 
agreement with the value found from the fit (solid line in Fig. \ref{refref}%
) for $H_{c2}\left( 0\right) =5\cdot 10^6\ G$, $Gi=0.01$, $f_{th}\sim 
10^{10}\ Hz$, and $\gamma \simeq 4$. 
 
To further confirm that in our $YBCO$ films the most probable physical 
mechanism for the onset of irreversibility is a dynamic crossover from 
unpinned to pinned vortex liquid we discuss now the frequency dependence of $%
T_{irr}$. Equation \ref{Tk} has a clear prediction for the frequency 
dependence of $T_{irr}$. To see it directly we may simplify it by using the 
experimentally determined smallness of value of the fit parameter $A\simeq 
0.0005$, which allows to expand Eq. \ref{Tk} (for not too large fields) as  
\begin{equation} 
T_k\approx T_c\left( 1-\frac 1\gamma \left( \frac{4Gi}{H_{c2}\left( 0\right)  
}\right) ^{\frac 12}\ln \left( f_{th}/f\right) \sqrt{\varepsilon _\theta B}%
\right)  \label{Tks} 
\end{equation} 
which results in a linear dependence of $T_{irr}$ upon $\ln (f)$ and a slope  
$S\equiv \partial Tirr/\partial \ln (f)\approx \frac{T_c}\gamma \left( \frac{%
4Gi}{H_{c2}\left( 0\right) }\right) ^{\frac 12}\sqrt{\varepsilon _\theta B}%
=T_cA\sqrt{\varepsilon _\theta B}\ln \left( f/f_{th}\right) $. Note that the 
slope is proportional to $\sqrt{B}$. This is indeed confirmed by the 
experimental data, as is demonstrated by the solid lines in Fig. \ref{reffr}%
. From this fit we get $S/\sqrt{B}=0.004$ and we can independently verify 
the parameter $A$ appeared in Eq. \ref{Tk} $A=S/\left( T_c\sqrt{\varepsilon 
_\theta B}\ln \left( f/f_{th}\right) \right) =0.0008$, which is in an 
agreement with the value obtained above. 
 
We note that the approximated expression for the frequency dependence of $%
T_{irr}$, Eq. \ref{Tks}, is valid in the whole experimentally accessible 
range of magnetic field since Eq. \ref{Tk} predicts a maximum in the slope $%
S $ at $B_{\max }=$ $\left( A\sqrt{\varepsilon _\theta }\right) ^{-2}\approx 
400\ Tesla$ for the experimental parameters. This value is, of course, 
beyond the experimental limits and, probably, even exceeds $H_{c2}$. 
 
Another support for the onset of the irreversibility in a vortex liquid is the $%
ac$ field amplitude dependence of the IRL. In both, thermal-activated (TAFF)
and pure flux-flow (FF) 
regimes the I-V curves are linear and the onset of the third harmonic is 
due to a change in the slope (from $\rho _{FF}$ to $\rho _{TAFF}$). In this case we expect the amplitude 
dependence for this onset. Contrary, at the melting transition the onset of 
irreversibility is sharp and is not expected to depend upon the amplitute of 
the $ac$ field. In our experiments we find a pronounced amplitude dependence 
of the IRL, thus confirming the above scenario. 
 
\subsection{Irradiated YBCO films} 
 
For the irradiated films the situation is quite different. The models for $%
T_{irr}\left( \theta \right) $ in unirradiated films cannot explain the 
experimental features exhibited in Figs. \ref{refuir} and \ref{refcir}, in 
particular the increase in $T_{irr}$ in the vicinity of $\theta =0$. Such 
discrepancy can only be due to the angular anisotropy introduced by columnar 
defects, i.e., the angle-dependent pinning strength. It was shown, both 
theoretically \cite{blatter,fisher2} and experimentally \cite{krusin}, that 
for magnetic field oriented along the defects the irreversibility line is 
shifted upward with respect to the unirradiated system. Thus, our results in 
Fig. \ref{refuir} suggest that the measured $T_{irr}\left( \theta \right) $ 
is a superposition of the angular variation of $T_{irr}$ in unirradiated 
film (denoted in this section as $T_{irr}^{REF}$) and the anisotropic 
enhancement of the pinning strength due to irradiation. 
 
We can estimate the latter contribution by employing the concept of a 
''trapping angle'' $\theta _t$, the angle between the external field and the 
defects at which vortices start to be partially trapped by columnar tracks. 
(For a schematic description, see Fig. $43$ in Blatter et al. \cite{blatter}%
). As we show in Appendix A,  
\begin{equation} 
\tan \left( \theta _t\right) \approx \sqrt{2\varepsilon _r/\varepsilon _l} 
\label{tantr} 
\end{equation} 
where $\varepsilon _r\left( T\right) $ is the trapping potential of a 
columnar defect and $\varepsilon _l$ is the line tension. In the experiment 
we cool down at a fixed $\theta $, and the onset of irreversibility must 
occur when $\theta =\theta _t\left( T\right) $, provided that the 
temperature is still larger than $T_{irr}^{REF}\left( \theta \right) $. 
Otherwise, the onset occurs at $T_{irr}^{REF}$. This defines the condition 
for the irreversibility temperature $T_{irr}$ for angles $\theta \leq \theta 
_c\equiv \theta _t\left( T_{irr}^{REF}\left( \theta _t\right) \right) \simeq 
50^o$ in our case:  
\begin{equation} 
\tan \left( \theta \right) =\tan \left( \theta _t\left( T_{irr}\right) 
\right) \approx \sqrt{2\varepsilon _r/\varepsilon _l}  \label{trapangle} 
\end{equation} 
At high-temperatures $\varepsilon _r\left( T\right) \propto \exp \left( -T/%
\widetilde{T}_{dp}^r\right) $, where $\widetilde{T}_{dp}^r$ is the depinning 
energy \cite{blatter}. Thus, we can write for $T_{irr}$:  
\begin{equation} 
T_{irr}\left( \theta \right) =\left\{  
\begin{array}{ll} 
T_{irr}^{REF}\left( \theta \right) -D\ln \left( C\left| \tan \left( \theta 
\right) \right| \right) & \theta \leq \theta _c \\  
T_{irr}^{REF}\left( \theta \right) & \theta >\theta _c 
\end{array} 
\right.  \label{tirrcd} 
\end{equation} 
where $D$ and $C$ are constants. This expression is in an agreement with our 
results shown in Fig. \ref{refuir} (solid line). We note, however, some 
discrepancy in the vicinity of $\theta =0$, where we find quite weak 
dependence of $T_{irr}$ on angle. We explain this deviation by considering 
the influence of relaxation, which, in the case of parallel defects depends 
on angle. The relaxation rate is maximal, when vortices are aligned along 
the defects and retains its normal ''background'' value for perpendicular 
direction \cite{anisotropicS}. Vortex, captured by a defect, can nucleate a 
double kink which slides out resulting in a displacement of a vortex on a 
neighboring column. In our irradiated samples the defect lattice is very 
dense (the matching field $B_\phi =4\;Tesla$, i.e., distance between columns  
$d\approx 220\;\AA $) and such double-kink nucleation an easy process. Thus, 
the irreversibility temperature should be shifted down around $\theta =0$ as 
compared to the ''ideal'', non-relaxed value, Eq. \ref{tirrcd}. This 
explains the reduction in $T_{irr}$ in Fig. \ref{refuir}. 
 
We may now conclude that in irradiated films, for angles less than the 
critical angle $\theta _c$, the irreversibility line is determined by the  
{\bf trapping angle }$\theta _t${\bf . }The Bose-glass transition can 
probably only be found for small angles within the lock-in angle $\theta 
_L\leq 10^o$. This conclusion is also indirectly confirmed in \cite{samoilov}%
. 
 
As was pointed out in the introduction, crossed defects should hinder the 
relaxation due to forced entanglement of vortices. Thus, the irreversibility 
temperature is expected to be closer to that predicted by Eq. \ref{tirrcd}. 
Fig. \ref{refcir} shows a good agreement of the experimental data with Eq.  
\ref{tirrcd} (solid line). To explain why defects crossed at large angle act 
collectively and force unidirectional magnetic anisotropy, we follow here 
the approach outlined in \cite{schuster}, and extend that description to 
account for arbitrary orientation of the external field with respect to the 
crossed columnar defects and to the $c-$axis. In Ref. \cite{schuster} the 
authors consider possible motion of vortices in a ''forest'' of crossed 
defects for field oriented along the $c-$axis. In our case of a dense 
lattice we may exclude from consideration free kink sliding and consider 
only depinning from the intersections. We sketch in Fig. \ref{forest} the 
two limiting situations: $(a)$ the external field is parallel to one 
subsystem of the columnar defects $\left( \theta =45^o\right) $ and $(b)$ 
the external field is oriented along the $c-$axis, between crossed columns $%
\left( \theta =0\right) $. In case $(a)$, Fig. \ref{forest}a, vortices can 
depin just by nucleation the single kinks which are sliding from 
intersection to intersection, or, by nucleation of super-kinks resulting in 
a kind of motion, similar to a variable-range hopping. This type of 
thermally assisted vortex depinning does not cost any additional energy on 
vortex bending. Another situation arises for field along the $c-$axis, Fig.  
\ref{forest}b. Now vortices can depin {\em only} via nucleation of multiple 
half-loops, which characteristic size depends upon current density. This 
results in additional barrier for vortex depinning, which even diverges at 
zero current \cite{blatter}. As a result, the relaxation rate is 
anisotropic, i.e, it is suppressed when the external field is oriented along 
the mid direction between the two subsystems of the crossed columnar 
defects. This is just opposite to a situation in uniformly irradiated 
samples. 
 
\section{Conclusions} 
 
We presented angle-resolved measurements of the irreversibility temperature 
in unirradiated $YBa_2Cu_3O_{7-\delta }$ film and in two films with columnar 
defects, induced by $6\ GeV$ Pb-ions irradiation, either parallel to the $c-$%
axis or 'crossed' in $\theta =\pm 45^o$. We find that in the unirradiated 
film the transition from irreversible to reversible state occurs {\bf above 
the melting line} and marks the {\bf crossover from pinned to unpinned 
vortex liquid}. In irradiated films, within the critical angle $\theta 
_c\simeq 50^o$, the irreversibility line is determined by the temperature 
dependent {\bf trapping angle}. For larger angles $T_{irr}$ is determined by 
the intrinsic anisotropy via the effective field. The formulae for $%
T_{irr}\left( \theta \right) $ for both unirradiated and irradiated films 
are given. We also discuss the possible influence of anisotropic enhancement 
in relaxation rate which leads to a smearing of the expected cusp at $\theta 
=0$ in the $T_{irr}\left( \theta \right) $ curve in the uniformly irradiated 
film. Finally, we demonstrate the collective action of crossed columnar 
defects, which can lead to suppression of relaxation and enhancement of 
pinning strength along the mid direction. 
 
{\bf Acknowledgments}{\em . }We are grateful to V. Vinokur, L. Burlachkov 
and P. Kes for illuminating comments. We thank S. Bouffard for help with Pb 
irradiation at GANIL, Caen (France). This research was supported in part by 
the Israeli Science Foundation administered by the Israeli Academy of 
Science and Humanities and the Heinrich Hertz Minerva Center for High 
Temperature Superconductivity, and in part by the DG XII, Commission of the 
European Communities, and the Israeli Ministry of Science and the Arts 
(MOSA). A. S. and R. P. acknowledge a support from the Israeli-France 
co-operation program AFIRST. R. P. acknowledges a support from the
the Clore Foudations and the hospitality of LSI at 
Ecole Polytechnique during part of this research.

\newpage  
 
\section*{Appendix A} 
 
We describe here the derivation of our Eq. \ref{tantr}, which differs 
slightly from the analogous Eq. ($9.173$) of Blatter et al. \cite{blatter}. 
We derive it using exactly the same approach (and notions) as in \cite 
{blatter}, but, in view of the experimental situation, avoid assumption of 
small angles, which allows Blatter et al. to approximate $\tan \left( \theta 
\right) \approx \sin \left( \theta \right) \approx \theta $. In order to 
estimate the trapping angle one has to optimize the energy change due to the 
vortex trapping by columnar defects. This energy is written as \cite{blatter}%
: 
 
\begin{mathletters} 
\begin{equation} 
\varepsilon \left( r,\theta \right) =\varepsilon _l\left[ r+\left( 
d^2+\left( \frac d{\tan \left( \theta \right) }-r\right) ^2\right) -\frac d{%
\sin \left( \theta \right) }\right] -r\varepsilon _r  \eqnum{1A} 
\label{energy} 
\end{equation} 
where $r\left( \theta \right) $ is the length of the vortex segment, trapped 
by a defect; $d$ is the distance between the columns; $\varepsilon _l$ is 
the line tension; and $\varepsilon _r$ is the trapping potential of the 
defects. The variation of Eq. \ref{energy} with respect to $r$ at fixed a 
angle $\theta $ defines the angular dependence of $r\left( \theta \right) $. 
The trapping angle $\theta _t$ can be found by solving the equation $r\left( 
\theta _t\right) =0$. This results in  
\begin{equation} 
\tan \left( \theta _t\right) =\frac{\sqrt{\varepsilon _r\left( 2\varepsilon 
_l-\varepsilon _r\right) }}{\left( \varepsilon _l-\varepsilon _r\right) }  
\eqnum{2A}  \label{tantra} 
\end{equation} 
which, at sufficiently small $\varepsilon _r$, can be approximated as  
\end{mathletters} 
\begin{mathletters} 
\begin{equation} 
\tan \left( \theta _t\right) =\sqrt{\frac{2\varepsilon _r}{\varepsilon _l}}%
+O\left( \varepsilon _r^{3/2}\right)  \eqnum{3A}  \label{tantrsim} 
\end{equation} 
Apparently, at very small angles we recover the original result of Ref.\cite 
{blatter}. In the paper, for the sake of simplicity, we use expression Eq.  
\ref{tantrsim} instead of the full expression Eq. \ref{tantra}. However, as 
noted above we cannot limit ourselves bto small angles and, generally 
speaking, the trapping angle may be quite large ($\theta _t\approx 40^o$ in 
our case). The error due to use of Eq. \ref{tantrsim} can be etimated as 
follows: at $\theta \approx 40^o$ Eq. \ref{tantra} gives $\varepsilon 
_r/\varepsilon _l\simeq 0.24$, whereas Eq. \ref{tantrsim} gives $\varepsilon 
_r/\varepsilon _l\simeq 0.35$, which is suitable for our implication of Eq.  
\ref{tantrsim}, since we consider exponential decrease of $\varepsilon _r$. 
Also, as shown in \cite{blatter} in a system with anisotropy $\varepsilon $, 
the trapping angle is enlarged by a factor of $1/\varepsilon $. 
 
\begin{figure}[tbp] 
\caption{The third harmonic signal $V_3$ versus temperature during 
field-cooling at $1\ Tesla$ for sample {\em REF} at $\theta =0,\ 10^o,\ 
30^o,\ 40^o,\ 60^o,\ 80^o,\ 90^o$.} 
\label{RawData} 
\end{figure} 
 
\begin{figure}[tbp] 
\caption{The irreversibility temperature in the unirradiated sample {\em REF} 
at two values of the external field: $H=0.5$ and $1\ Tesla$. The solid lines 
are fits to Eq. \ref{Tk}.} 
\label{refref} 
\end{figure} 
 
\begin{figure}[tbp] 
\caption{The frequency dependence of $T_{irr}$ in the unirradiated sample  
{\em REF} at two values of the external field: $H=0.5$ and $1\ Tesla$. The 
solid lines are fits to Eq. \ref{Tks}.} 
\label{reffr} 
\end{figure} 
 
\begin{figure}[tbp] 
\caption{The irreversibility temperature for two samples: {\em REF} 
(unirradiated - open circles) and {\em UIR }(irradiated along the $c-$ axis 
sample, - filled circles) at $H=1\ Tesla$. Solid lines are fits to Eq. \ref 
{Tk} and Eq. \ref{tirrcd}, respectively.} 
\label{refuir} 
\end{figure} 
 
\begin{figure}[tbp] 
\caption{The irreversibility temperature for two samples: {\em REF} 
(unirradiated - open circles) and {\em CIR }(irradiated along $\theta =\pm 
45^o$ - filled circles) at $H=0.5\ Tesla$. Solid lines are fits to Eq. \ref 
{Tk} and Eq. \ref{tirrcd}, respectively.} 
\label{refcir} 
\end{figure} 
 
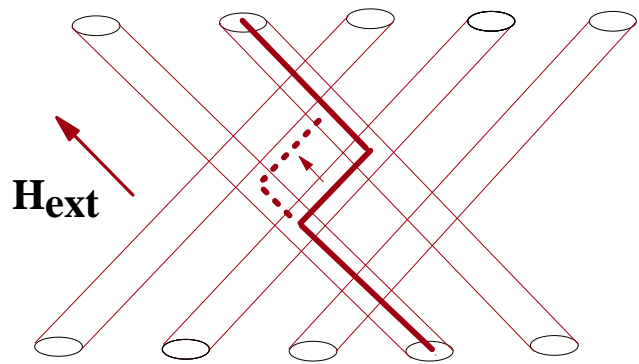
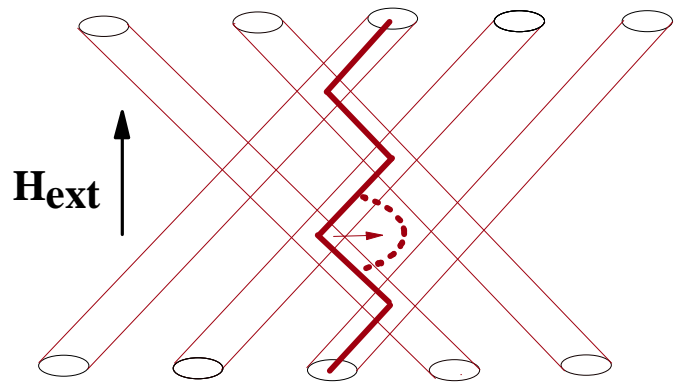
\begin{figure}[tbp] 
\caption{Schematic description of a possible depinning modes of a vortex 
line in the case of crossed columnar defects; - (a) magnetic field is 
directed along $\theta =45^o$; (b) magnetic field is along $\theta =0$.} 
\label{forest} 
\end{figure} 
\end{mathletters} 
 
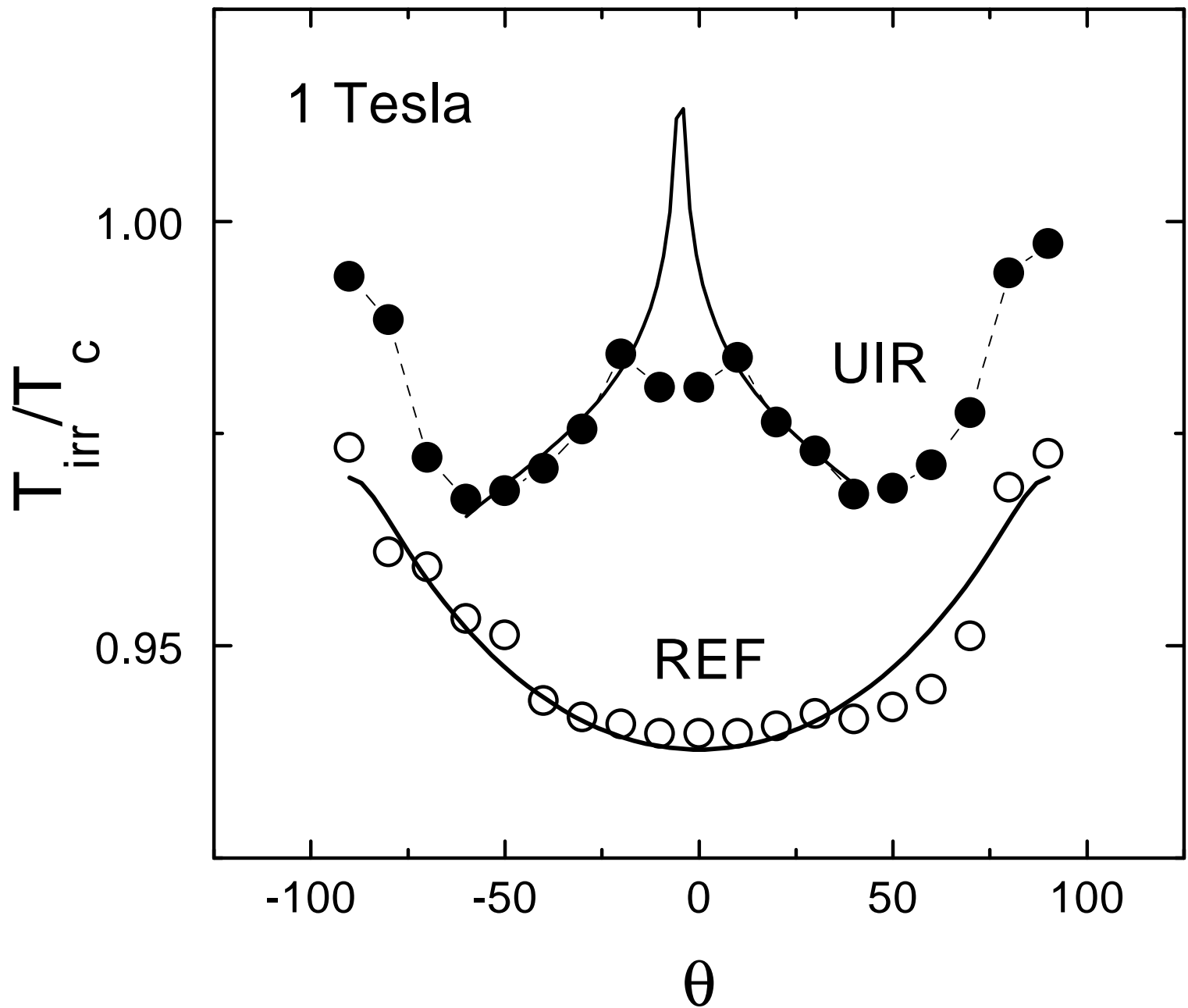
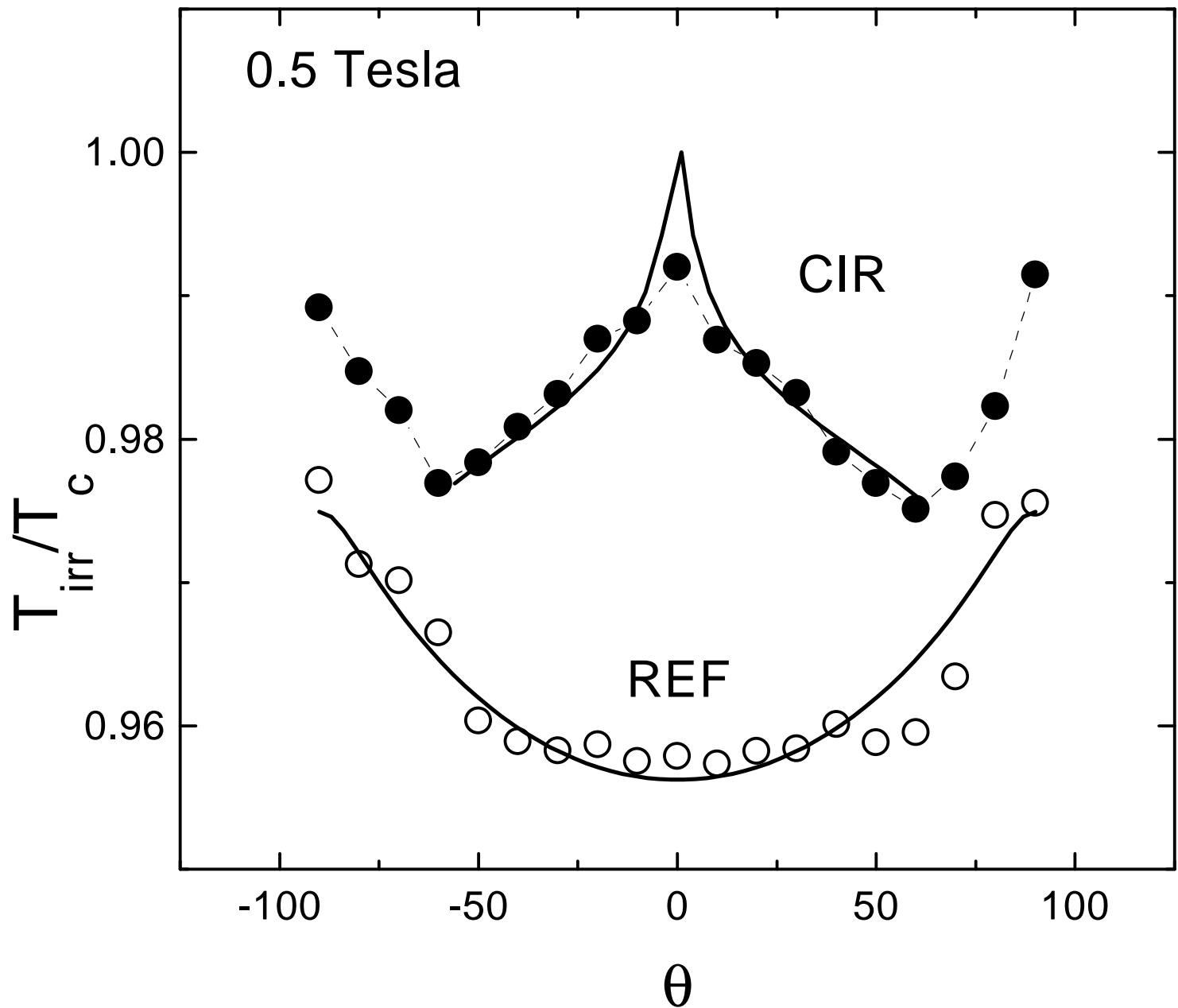
\end{document}